Experimental determination of phonon thermal conductivity and Lorenz ratio of single crystal metals: Al, Cu and Zn

Mengliang Yao[1], Mona Zebarjadi[2], and Cyril P. Opeil[1]
[1]*Department of Physics, Boston College, Chestnut Hill, Massachusetts 02467, USA*
[2]*Department of Electrical and Computer Engineering and Department of Materials Science and Engineering, University of Virginia, Charlottesville, Virginia 22904, USA*

We use a magnetothermal resistance method to measure lattice thermal conductivity of pure single crystal metals over a wide range of temperatures. Large transverse magnetic fields are applied to suppress electronic thermal conduction. The total thermal conductivity and the electrical conductivity are measured as functions of applied magnetic field. The lattice thermal conductivity is then extracted by extrapolating the thermal conductivity versus electrical conductivity curve at zero electrical conductivity. We used this method to experimentally measure the lattice thermal conductivity and Lorenz number in single crystal Al (100), Cu (100) and Zn (001) in a temperature range of 5 to 60 K. Our results show that the measured phonon thermal conductivity versus temperature plot has a peak around $\Theta_D/10$, and the Lorenz number is found to deviate from the Sommerfeld value in the intermediate temperature range.

# I. Introduction

Thermal transport in metals is complex due to the coexistence of electron and phonon conduction and the strong interaction between them. The total thermal conductivity ($\kappa_{tot}$) has two components: The electronic part ($\kappa_e$), and the phononic part ($\kappa_{ph}$). The total thermal conductivity is the sum of these two components $\kappa_{tot} = \kappa_e + \kappa_{ph}$. In high purity metals, such as single crystal metals, the electrons dominate thermal conduction due to their high density and the fact that the Fermi velocity of electrons is much greater than the speed of sound, *i.e.* the speed of phonons. [1] As a result, it is challenging to directly measure the phonon thermal conductivity of metals.

Because of the rarity of high magnetic fields ($H \geq$ 10 kOe) coupled with low-temperature measurements at that time, starting in the late 1950s, researchers measured the electrical resistivity of a series of dilute alloys at liquid He temperatures and extrapolated back to the conditions of a zero-impurity content in order to approximate the electronic thermal conductivity contributions in pure metals. [2] Determining $\kappa_e$ is made possible by applying the Wiedemann-Franz law where $\kappa_e/\sigma \propto T$. [5-7] However, this practice is both time consuming and laborious due to the number of samples and measurements needed rather than measuring only one single crystal.

Another possible way to probe the phonon thermal conductivity is by applying a large transverse magnetic field to a single specimen. By applying such a large field, one can separate the electronic and phononic part of thermal conductivity, assuming the transverse magnetic field eliminates virtually all electronic conduction. This technique has already been successfully applied to semiconductors [3] and semimetals.[4] Since carrier mobility is relatively large in semiconductors and semimetals, the magnetic field available in a modern laboratory ($\sim$10 T) is often large enough to saturate the magnetothermal resistance and isolate the phonon thermal conductivity. Assuming the electronic thermal conduction is virtually eliminated and the lattice thermal conduction is field independent, the total thermal conductivity measured under large magnetic field is equal to the lattice thermal conductivity. This strategy is limited to the low temperature regime ( $T < 100$ K ) because at higher temperatures ( $T > 100$ K ) extremely large magnetic fields ($\gg$ 10 T) are required to suppress of the electronic component of thermal conductivity. Such larger magnetic fields are not only impractical, but can also effect the lattice thermal conduction via electron-phonon interaction. This work focuses on single crystal metals in the low-temperature regime where metals have both a low electrical resistivity and large mobility values. At low temperatures, the thermal conduction of metals can be strongly affected by the magnetic fields available in the laboratory. [8]

The deflection angle $\gamma$ is defined as the deviation of an electron away from its previous linear motion under the influence of the applied magnetic field between collisions and is used to identify the strength of the field:[8]

$$\gamma = \omega_c \tau = \mu B = \frac{\sigma_0 B}{ne} \tag{1}$$

where, $\omega_c$ is the cyclotron frequency, $\tau$ is the relaxation time, $\mu$ is the mobility, $B$ is the magnetic induction strength, $\sigma_0$ is the zero field electrical conductivity, $n$ is the carrier concentration, and $e$ is the elementary charge. If $\gamma \gg 1$ then the field is said to be a classically large magnetic field. [9] In practice, even when $\gamma \sim 1$, a significant suppression of electronic conduction can be observed.[8-12] The magnetic field corresponding to $\gamma \sim 1$ is defined here as the threshold field $B_{th}$. As temperature decreases, the threshold field decreases due to the decline of sample electrical resistivity and nearly constant carrier concentration. We estimate the $\gamma$ values for Al, Cu and Zn at 10 T and 100 K are 0.05, 0.15 and 0.11, while the corresponding values at 50 K under the same applied field are 0.34, 1.01 and 0.39. Therefore, we find that when the temperature is below 100 K, the suppression of electronic thermal conductivity becomes noticeable and is enhanced as temperature decreases. The region $T < 100$ K corresponds to the area of highest mobility for Al, Cu and Zn.

The Lorenz ratio (or Lorenz number) is defined in zero field as: [22, 23, 26]

$$L \equiv \frac{\kappa_e}{\sigma_0 T} \tag{2}$$

and characterizes the transport properties of electrons. It is widely used to estimate the electronic contribution to the total thermal conductivity from the electrical resistivity measurements, [13-16] where $\kappa_e$ is the electronic thermal conductivity, $\sigma_0$ is the electrical conductivity, and $T$ is the absolute temperature. The Sommerfeld value

$$L_0 = \frac{\pi^2}{3}\left(\frac{k_B}{e}\right)^2 = 2.443 \times 10^{-8}\ \mathrm{V^2/K^2} \tag{3}$$

is derived from a consideration of the Wiedemann-Franz law in metals by using Somerfield expansion and neglecting the thermoelectric term.[1] In deriving the constant Lorenz ratio, it is assumed that all scattering is elastic, including the electron-phonon scattering at high temperatures and the impurity scattering at low temperatures, and the same relaxation time is used for both electrical and thermal conductivities.[2] However, in the intermediate temperature range, the above two conditions may fail and the Lorenz ratio can significantly deviate from $L_0$. In order to fully describe the temperature dependence of the Lorentz number, we introduce separate relaxation time considerations for the electronic thermal conductivity and electrical conductivity. By introducing different relaxation times for electrical transport and electronic thermal transport without electron-phonon Umklapp processes, the temperature dependence of the dimensionless Lorenz ratio of free electrons in zero field $\tilde{L} \equiv L/L_0 = \frac{\kappa_0 - \kappa_{ph}}{L_0 \sigma_0 T}$ can be described using the following formula: [2]

$$\tilde{L} = \frac{\beta + \left(\frac{T}{\Theta_D}\right)^5 J_5\left(\frac{\Theta_D}{T}\right)}{\beta + \left(\frac{T}{\Theta_D}\right)^5 J_5\left(\frac{\Theta_D}{T}\right)\left[1 + \frac{3\alpha^2}{\pi^2}\left(\frac{\Theta_D}{T}\right)^2 - \frac{1}{2\pi^2}\frac{J_7(\Theta_D/T)}{J_5(\Theta_D/T)}\right]} \tag{4}$$

where $\beta$ describes the purity of the sample, $\alpha$ is the ratio between Fermi wave vector $k_F$ and Debye wave vector $q_D$, $J_n$ is defined as

$$J_n\left(\frac{\Theta}{T}\right) \equiv \int_0^{\Theta/T} \frac{x^n e^x}{(e^x - 1)^2} dx \tag{5}$$

and $\Theta_D$ is the Debye temperature which can be estimated from the electrical resistivity through Bloch's $T^5$ law [2, 24]

$$\rho(T) = \rho_{imp} + \alpha_{el-ph}\left(\frac{T}{\Theta_R}\right)^5 J_5\left(\frac{\Theta_R}{T}\right) \tag{6}$$

where $\rho_{imp}$ is the residual electrical resistivity, $\alpha_{el-ph}$ is the electron-phonon coupling constant and $\Theta_R$ is the resistivity Debye temperature.

Only a few results have been reported on the phonon thermal conductivity and Lorenz ratio of single crystal metals [5-7,18-21,25]. These limited data are due to the fact that the experimental setup is not easy to realize in the laboratory, controlling a large field and temperature simultaneously. Also, there are restrictions on the materials that can be explored by the magnetothermal resistance method, because a material requires a sufficiently high mobility at low temperature so that electronic saturation can be achieved. This applies to metals and some small band gap semiconductors.

The advent of the Physical Properties Measurement System (PPMS) from Quantum Design makes the experimental measurement possible and reliable with the Thermal Transport Option (TTO). The aim of this paper is to present experimental techniques for determining phonon thermal conductivity and Lorenz ratio of single crystal metals at low temperatures. These results provide the basis for a comparison between experimental and future theoretical material calculations. Recently, the phonon thermal conductivity of several single crystal metals (Ni, Ag, and Au) have been calculated from first principles at a higher temperature regime than our experimental data. [27, 28]

## II. Experimental Details

Our samples were commercially obtained single crystal metals: Cu (100) 99.99% from MTI Co.①, while Al (100) 99.999% and Zn (001) 99.999% Goodfellow Cambridge Ltd. Samples were cut into typical dimensions of 1 × 2 × 10 $mm^3$, and Ag contacts were sputtered onto the surface to provide low resistance solder contacts for a standard 4-probe resistance measurement. Our sample dimensions allow resistivity measurements to be made with current and voltage along the same axis. This is referred to as the *xx*-configuration. The 4-probe resistivity or Hall measurements with I and V in orthogonal directions is referred to as the *yx*-configuration. Reduced sample resistivity at low temperatures makes it impossible to measure in the *yx*-configuration.

Thermal conductivity measurements were performed using the thermal transport option (TTO) of the PPMS, where the heat flow in the sample is perpendicular to the applied magnetic field (*i.e.* transverse orientation). The measurements were performed in temperature range of 60 K to 5 K and magnetic field range of $0 \sim 90$ kOe. Resistivity data was obtained using a LR-700 AC resistance bridge from Linear Research Inc. by fixing temperature and slowly sweeping the applied magnetic field. Our propagated measurement error was calculated from a consideration of standard deviation and determined to be 8% for $\kappa_{ph}$ and 14% for the Lorenz ratio.

## III. Analysis of Magneto-transport Measurements

Residual resistance ratio (RRR), defined by $\rho_{300\mathrm{K}}/\rho_{2\mathrm{K}}$, and resistivity Debye temperature $\Theta_R$, are fitted using Eq. (6) and summarized in Table 1. Our Debye temperature $\Theta_D$ from the heat capacity measurements of the same samples confirms the consistency with $\Theta_R$.

**Table 1: Summary of magnetoresistance measurements**

| Specimen | RRR | $\Theta_R/\mathrm{K}$ | $\Theta_D/\mathrm{K}$ | Field Direction | Crystal Structure |
|---|---|---|---|---|---|
| Al (100) | 200 | 415 | 395 | (100) | fcc |
| Cu (100) | 100 | 337 | 323 | (100) | fcc |
| Zn (001) | 800 | 221 | 233 | (001) | hex |

Generally, transport measurements made in zero or low magnetic fields are considered as scalar quantities. However, as magnetic field effects increase, these quantities must be treated as tensor quantities. These transport coefficient tensors exhibit non-zero off diagonal elements due to Hall effects of the applied field. Because the electrical transport coefficients are tensors, the electrical conductivity, $\sigma_{xx}$is not in general equal to the inverse of its corresponding electrical resistivity component, *i.e.* $\sigma_{xx} \neq 1/\rho_{xx}$, except in zero field. Electrical conductivity $\sigma_{xx}$ can be derived from magnetoresistance $\rho_{xx}/\rho_0$ using Eq. (7) and Eq. (8). These formulae require a fitting of the magneto-resistance data in order to derive the two fitting parameters, namely $\gamma = \mu B$ and $c$ to calculate the electrical conductivity, $\sigma_{xx}$, where the formulae are derived for a transverse magnetoresistance of a square Fermi surface with constant relaxation time [8]:

$$\frac{\rho_{xx}}{\rho_0} = \frac{\left[\frac{4c}{\pi}\left(1-\frac{2\gamma}{\pi}\tanh\frac{\pi}{2\gamma}\right)+\frac{1}{1+\gamma^2}\right]\left(\frac{4c}{\pi}+1\right)}{\left[\frac{4c}{\pi}\left(1-\frac{2\gamma}{\pi}\tanh\frac{\pi}{2\gamma}\right)+\frac{1}{1+\gamma^2}\right]^2+\left[\frac{8\gamma c}{\pi^2}\left(1-\mathrm{sech}\frac{\pi}{2\gamma}\right)-\frac{\gamma}{1+\gamma^2}\right]^2} \tag{7}$$

① MTI Corporation

$$\sigma_{xx} = \frac{\frac{4c}{\pi}\left(1 - \frac{2\gamma}{\pi}\tanh\frac{\pi}{2\gamma}\right) + \frac{1}{1+\gamma^2}}{\left(\frac{4c}{\pi}+1\right)\rho_0}$$

(8)

where $\gamma$ and $c$ are the dimensionless fitting parameters.

In order to extract $\kappa_{ph}$ from $\kappa_{tot}$, the functional form of both $\kappa_{xx}(B)$ and $\sigma_{xx}(B)$, the following formulae are required, which are derived for isotropic single band materials:

$$\sigma(B) = \frac{\sigma_0}{1+\mu_d^2 B^2}$$

(9)

$$\kappa_{tot}(B) = \kappa_{ph} + \frac{\kappa_0}{1+\mu_t^2 B^2}$$

(10)

where $\sigma_0$ and $\kappa_0$ are the measurements in zero field, $\mu_d$ and $\mu_t$ are drift and thermal mobility values, respectively. Combining (9) and (10) together:

$$\kappa_{tot}(\sigma) = \kappa_{ph} + \frac{\kappa_0 - \kappa_{ph}}{1+\lambda^2\left(\frac{\sigma_0}{\sigma}-1\right)}$$

(11)

where $\lambda = \mu_t/\mu_d$ is considered as a fitting parameter. $\kappa_{tot}(\sigma)$ represents the relationship between $\sigma$ and $\kappa_{tot}$. If $\lambda = 1$, then $\kappa_{tot}$ and $\sigma$ are linearly related, which is observed at high temperatures [3]. However, in the low temperature regime, $\lambda \neq 1$ and a non-linear relationship is observed between $\kappa_{tot}$ and $\sigma$.

Fig. 1 shows the temperature dependent behavior of the total thermal conductivity and electrical resistivity under various applied magnetic fields. As mentioned previously, the threshold field, $B_{th}$, decreases as temperature is reduced. Therefore, when applying the maximum field of 90 kOe, we observe a significant suppression of the electronic transport below 60 K. Furthermore, we note the temperature dependent electrical resistivity of Zn in the field has minima unlike the behavior of the Al and Cu samples in the same temperature region. This behavior is due to the fact that Zn is a compensated metal, [8] which results in a large magnetoresistance phenomenon at low temperatures.

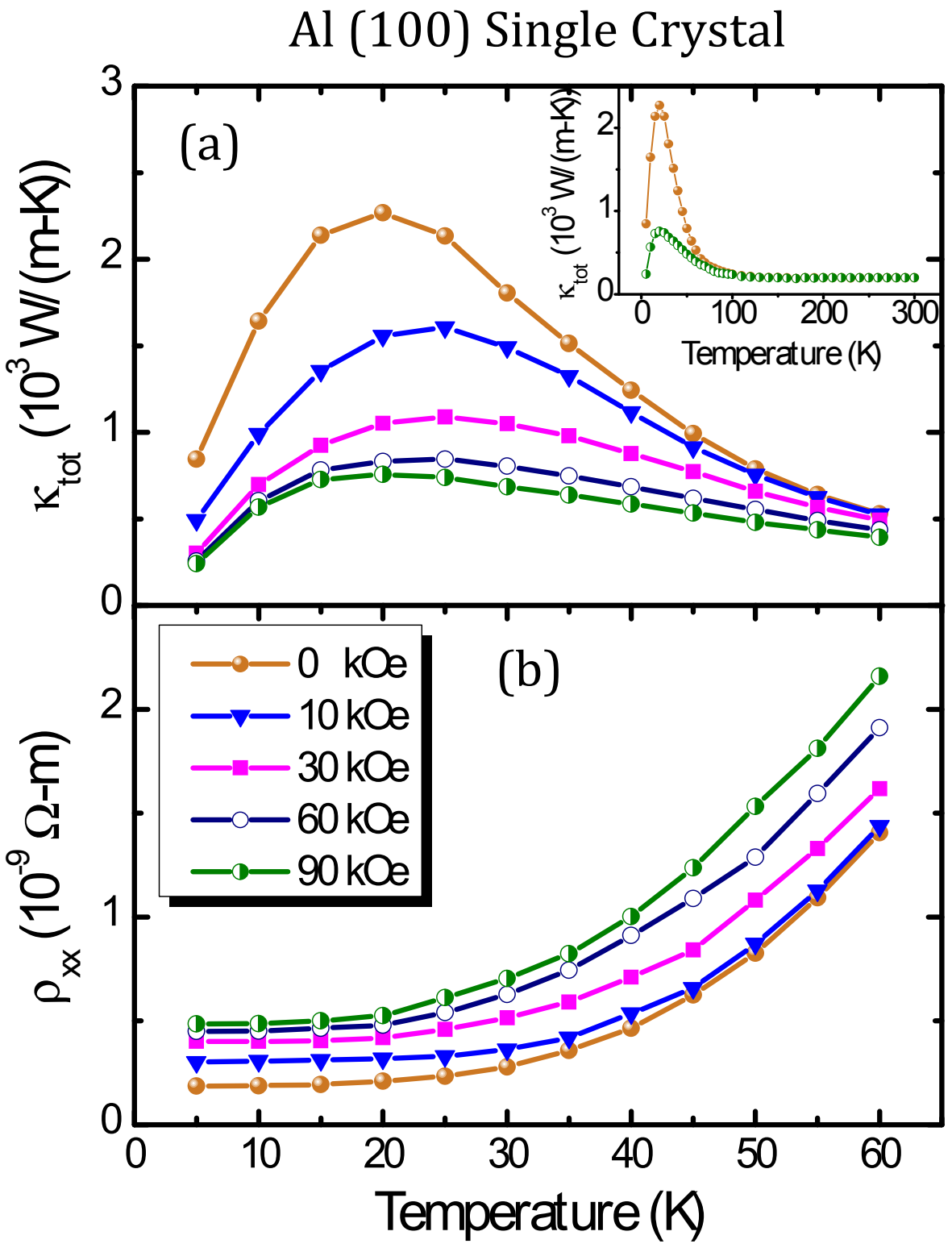

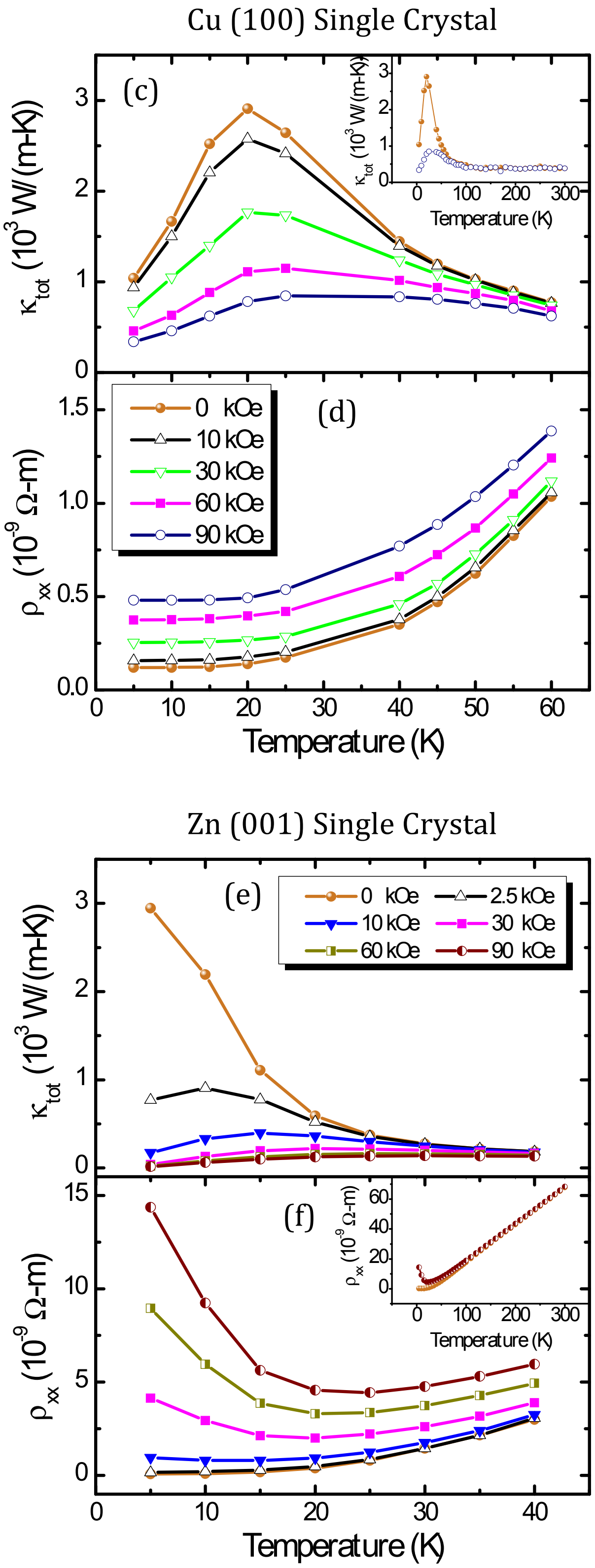

**FIG. 1: Thermal conductivity 1(a), 1(c), 1(e) and electrical resistivity 1(b), 1(d), 1(f) are plotted against temperature between 5 K and 60 K (for Al and Cu) or 40 K (for Zn) in different magnetic fields. The inset for 1a, 1c and 1f illustrate the magnetic field effect between 0 and 90 kOe on thermal conductivity for Al and Cu, electrical resistivity for Zn, as a function of temperature.**

In order to extract $\kappa_{ph}$ from $\kappa_{tot}$, one needs to plot the measured thermal conductivity vs. the calculated electrical conductivity, see Eq. (7), (8) and (11). Fig. 2 shows $\kappa_{tot} - \sigma$ curves at constant temperature (5 K and 40 K) for each metal from which we can extrapolate $\kappa_{ph}$. At high temperature, *e.g.* $T > 100$ K we expect the $\kappa_{tot} - \sigma$ curves to have a linear relationship, since $\lambda = 1$. As temperature decreases one expects $\kappa_{tot} - \sigma$ curve to deviate from a linear relationship, because the electrical resistivity and thermal conductivity have differing responses to the applied field. We find evidence for this behavior in the $\kappa_{tot} - \sigma$ curves in Fig. 2. These differing responses result from the different relaxation times for $\sigma$ and $\kappa$ in the field. Fig. 2 confirms this expected behavior, as noted by comparing the linearity of the curves at 5 K and 40 K. The intercepts of the curves in Figs. 2(a) – 2(f) represent the value of $\kappa_{ph}$, which is assumed to be field independent and is a condition of our analysis. In the upper left inset Figs. 2(a)(i), 2(c)(i), 2(e)(i) for 5 K we demonstrate the fit of magnetoresistance according to eqn. (7). The upper left inset at 40 K Figs. 2(b)(i), 2(d)(i), 2(f)(i) show the derived field dependence of electrical conductivity from the magnetoresistance fit to theory. These curves confirm the validity of eqn. (7) as a fit of magnetoresistance data to theory (red fit line). The field dependence of total thermal conductivity is taken directly from our thermal transport measurements, and is shown in the lower right inset Figs. 2(b)(ii), 2(d)(ii), 2(f)(ii) for the 40 K data. The bottom right inset Figs. 2(a)(ii), 2(c)(ii), 2(e)(ii) at 5 K figures show an expanded region of $\kappa_{tot}\sim\sigma$ curves from 50 kOe to 90 kOe. These curves extrapolate the sample behavior to the intercepts corresponding to the high field limit and confirm a high correlation between theory and measurement.

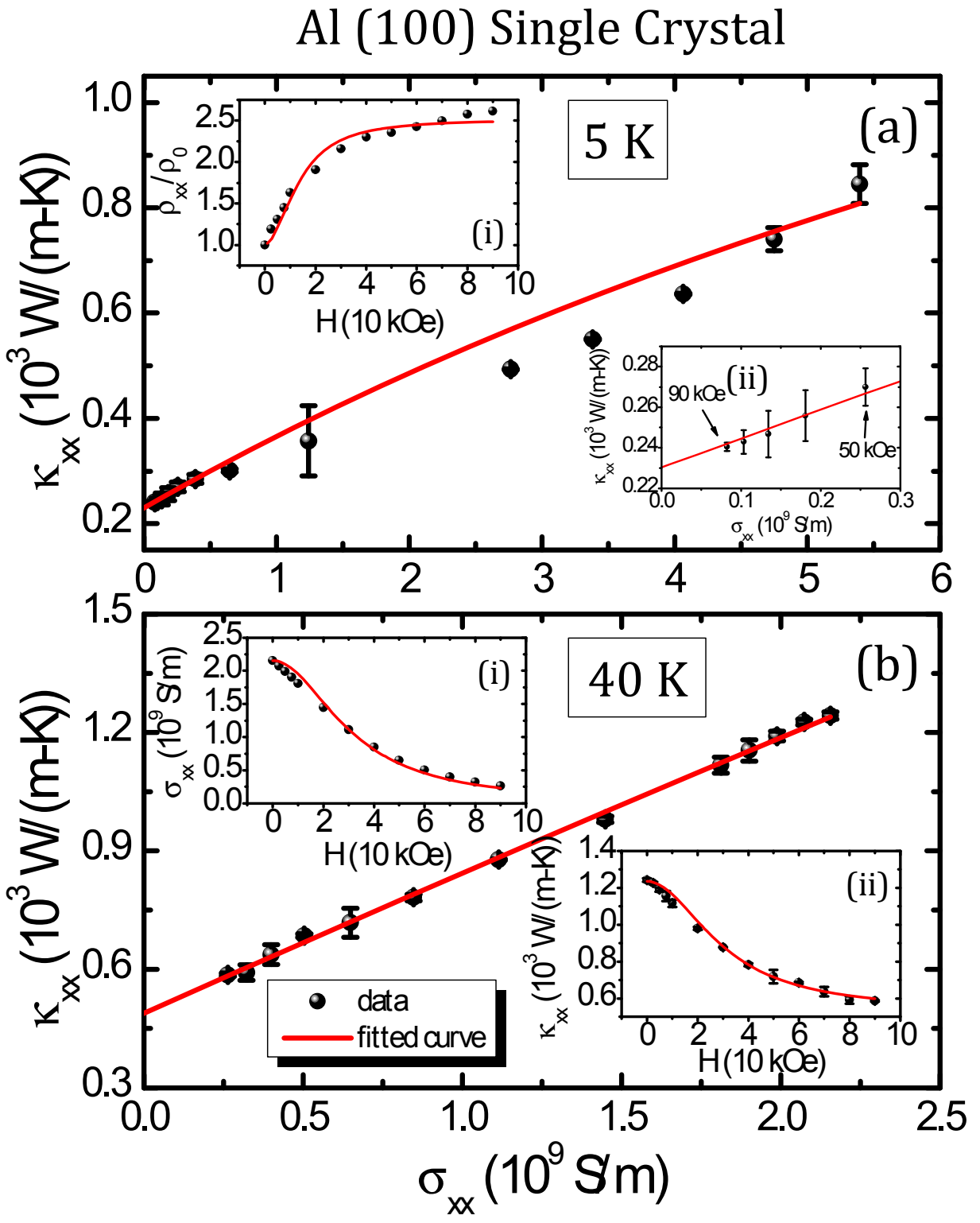

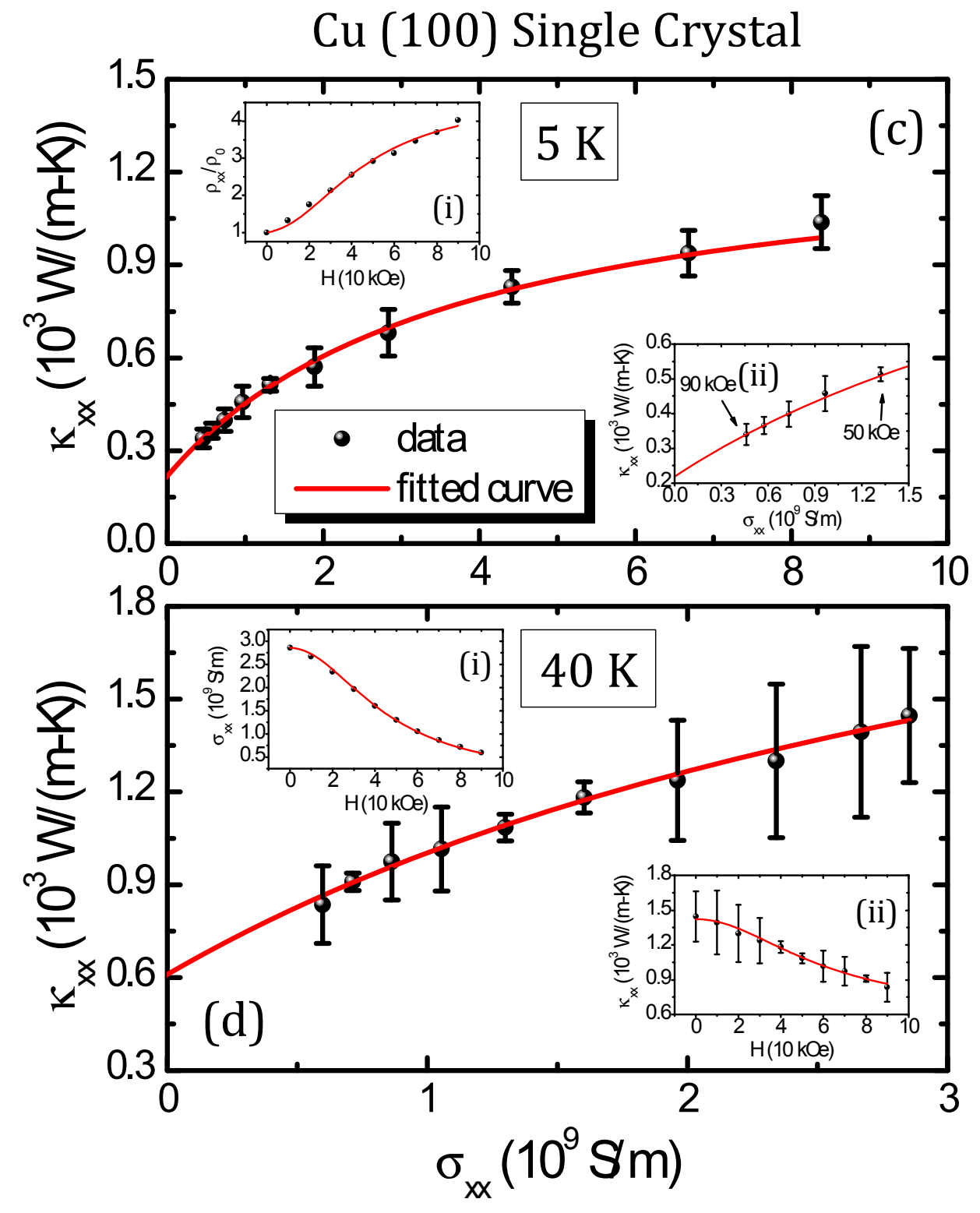

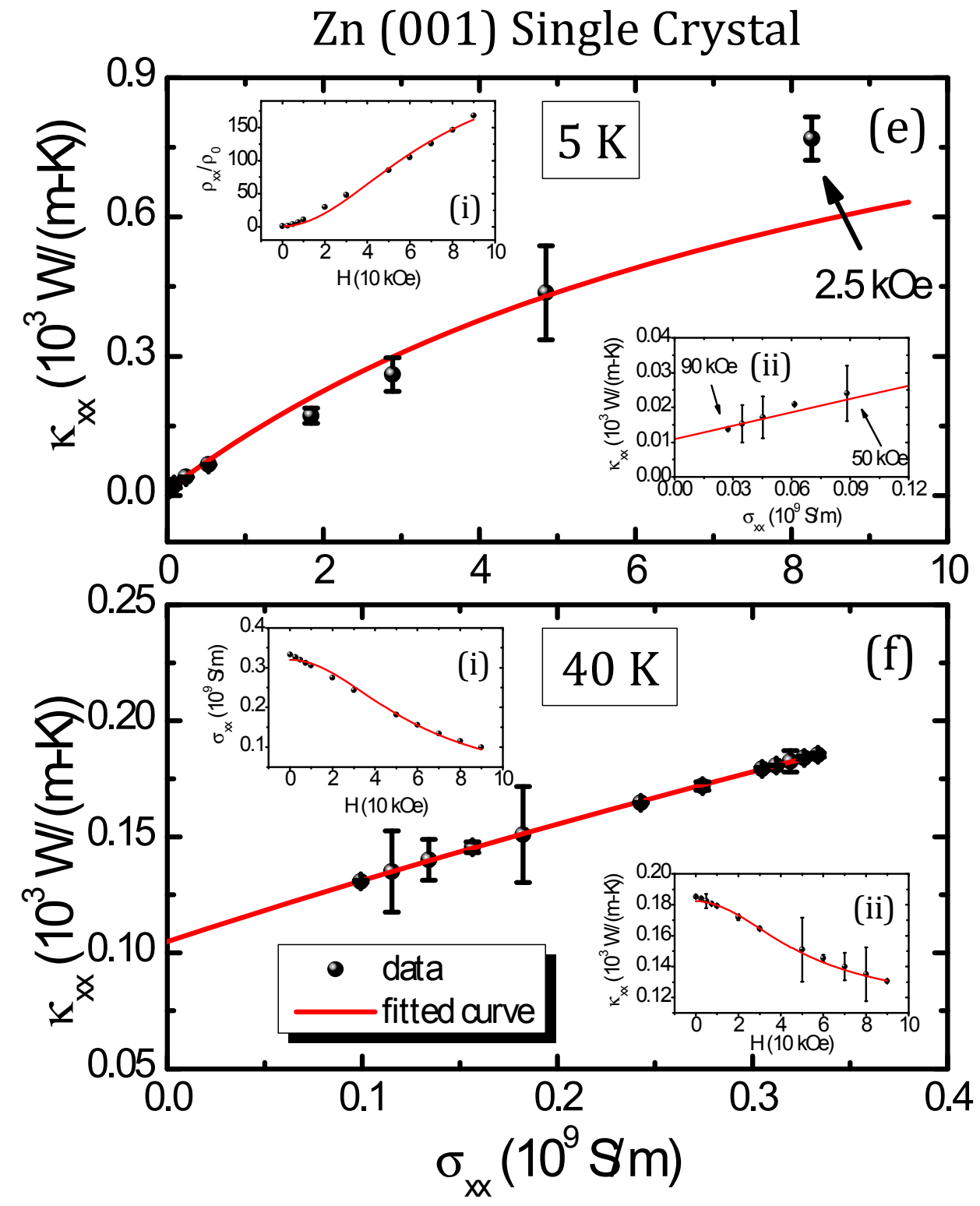

**FIG. 2: Total thermal conductivity is plotted against**

**electrical conductivity at 5 K and 40 K with a fit to eqn. (11). The top left inset in 5 K figure shows the magnetoresistance curve vs. the field along with a fit eqn. (7). The bottom right inset to Figs. 2(a)(ii), 2(c)(ii), 2(e)(ii) show expanded region of the curve at high fields. Both insets for the data at 40 K Figs. 2(b), 2(d), 2(f) show the field behavior of electrical and thermal conductivities along with their fit to theory eqns. (9), (10).**

Fig. 3 shows temperature-dependent phonon thermal conductivity and dimensionless Lorenz ratio for Al, Cu and Zn. Also, Fig. 3 demonstrates how large the phonon thermal conductivity $\kappa_{ph}$ is in comparison to the total thermal conductivity $\kappa_{tot}$. In Fig. 3(a), 3(c), 3(e), $\kappa_{ph}$ and $\kappa_{tot}(90\,\mathrm{kOe})$ curves are plotted vs. temperature; they demonstrate the effective suppression of thermal conductivity with magnetic field (90 kOe). We observe the $\kappa_e$ suppression is greatest at low temperature as expected because the threshold field, $B_{th}$, is minimum at low temperatures. The difference between the two curves is smallest at the lowest temperature, confirming that the classically large magnetic field is reached for low temperatures. It is noted that at higher temperatures, *e.g.* 60 K, a proportionately larger magnetic field is necessary to suppress the electron contribution to thermal conduction. The insets to Fig. 3(a), 3(c), 3(e) show the phonon contribution of thermal conductivity to the total thermal transport. The calculations demonstrate the ratio $\kappa_{ph}/\kappa_{tot}(0\,\mathrm{kOe})$ increases by up to 50% in the experimental temperature range investigated for these electron dominated metals (discussed further in Sect. IV). Because the primary carriers of thermal energy in these metals are electrons, both the electronic and total thermal conductivity should have a similar temperature dependence. This behavior is illustrated in the insets to Figs. 3(b), 3(d), 3(f). The dimensionless Lorenz ratio is shown in Figs. 3(b), 3(d), 3(f) and indicates a deviation from unity as temperature increases. Berman and MacDonald have shown the deviation from the Sommerfeld value in 1950s, although they calculated their Lorenz ratio from the total thermal conductivity $\kappa_{tot}$ rather than the electronic thermal conductivity $\kappa_e$.[29] The direct cause of this deviation is the inelastic scattering between electrons and phonons and a difference in the relaxation time required for thermal transport and electrical processes during the electron-phonon scattering. The relaxation time for the thermal transport process is shorter than the one for the electrical process. The behavior of the temperature dependent dimensionless Lorenz ratio for metals is well described by eqn. (4), as shown by our fit to theory. Following the trend of the fitting function eqn. (4), we expect the recovery of the W-F law and the Sommerfeld value $L_0$ to be restored at room temperature.

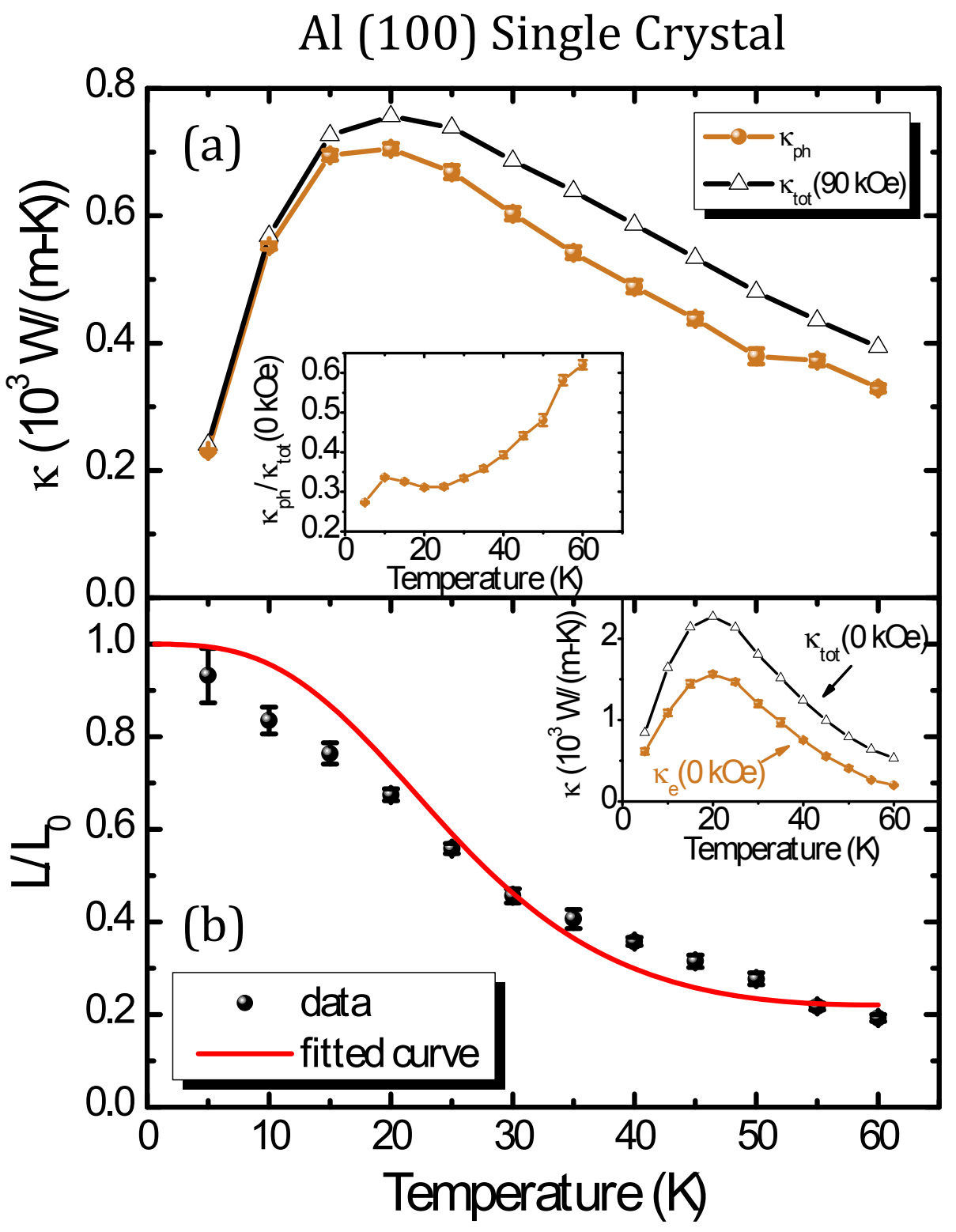

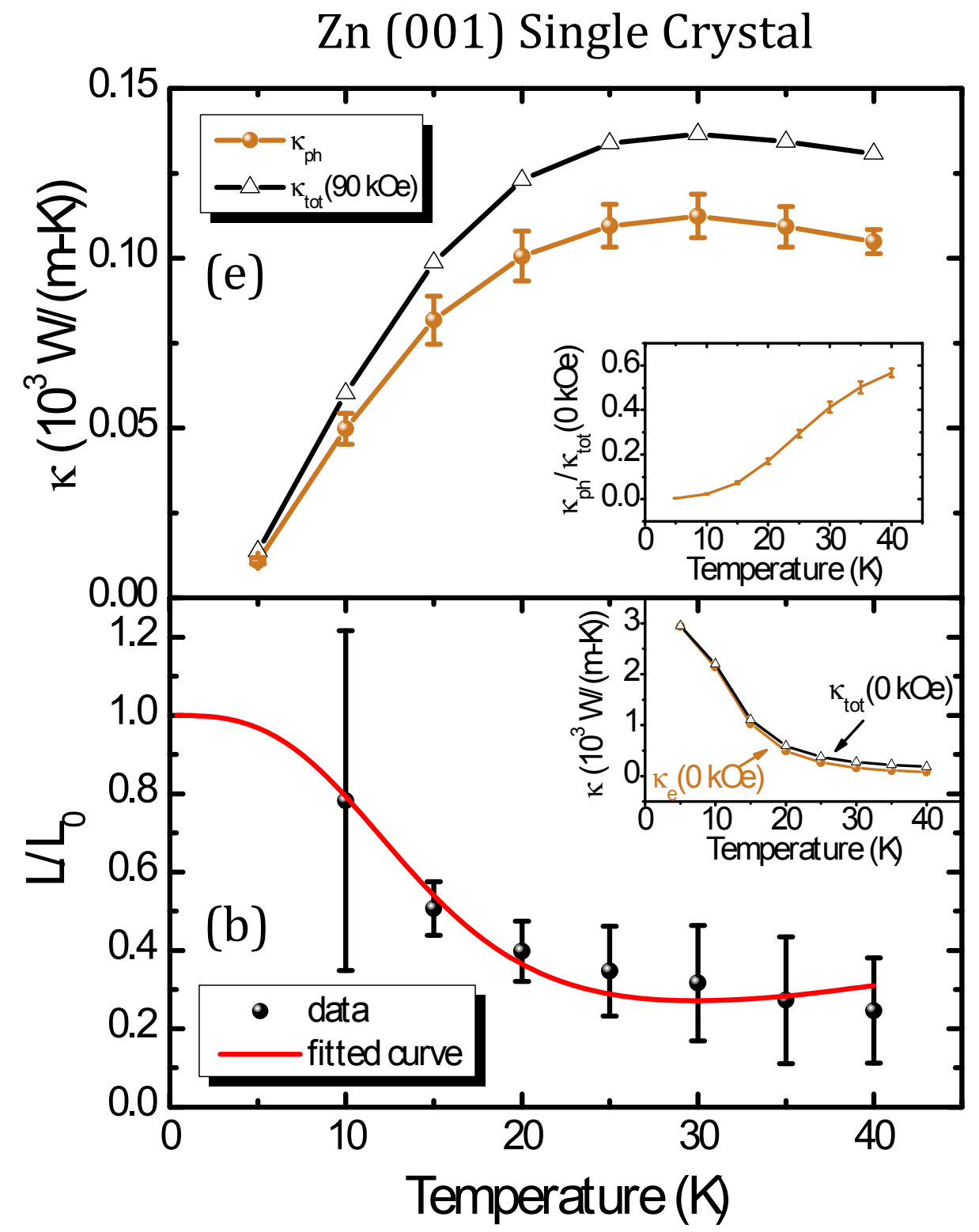

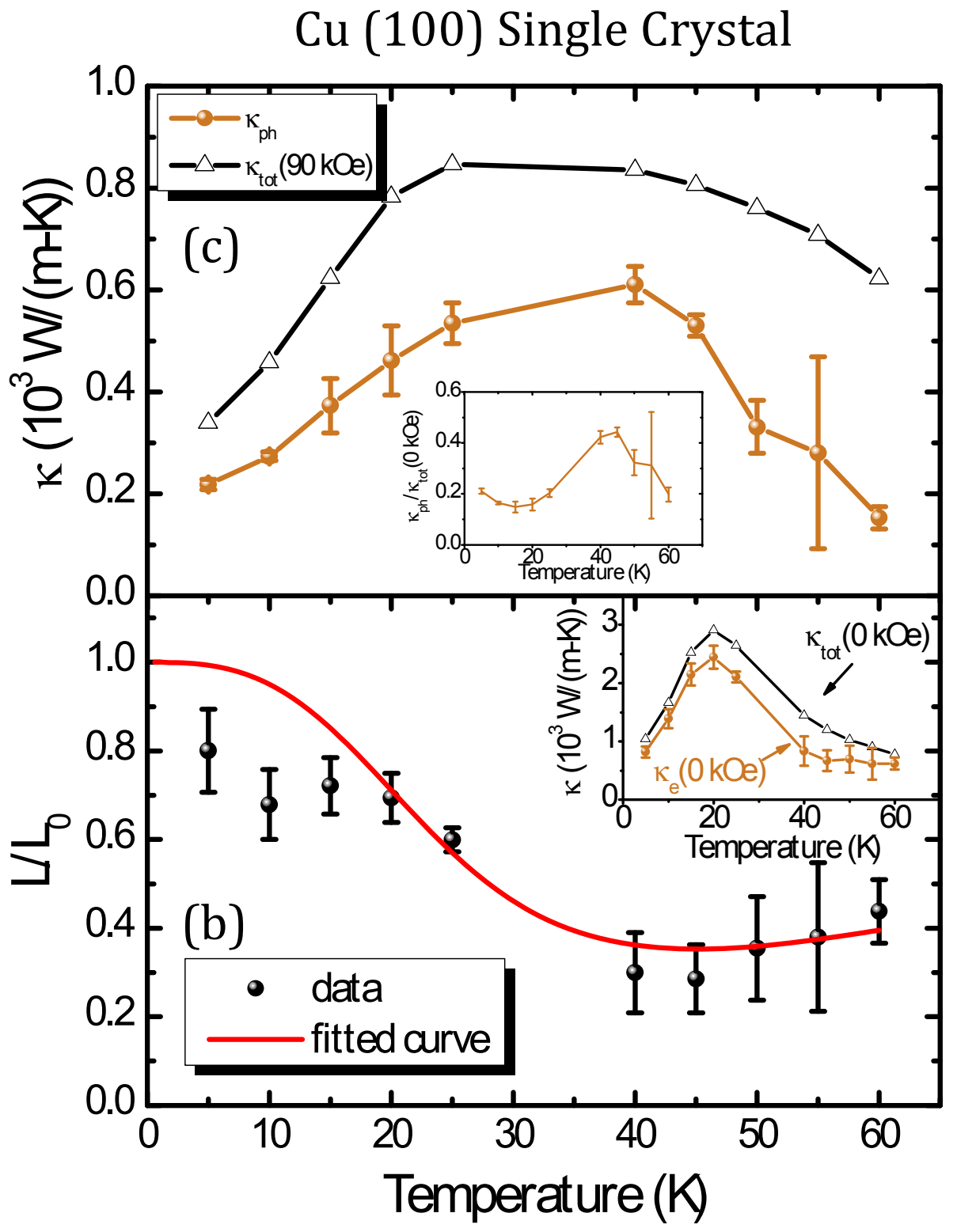

**FIG. 3: Phonon thermal conductivity Figs. 3(a), 3(c), 3(e) and dimensionless Lorenz ratio Figs. 3(b), 3(d), 3(f) are plotted against temperature for Al, Cu and Zn single crystals. Figs. 3(a), 3(c), 3(e) also show a comparison between $\boldsymbol{\kappa_{ph}}$ and $\boldsymbol{\kappa_{tot}(90\,\mathrm{kOe})}$; the insets to them show the ratio $\boldsymbol{\kappa_{ph}/\kappa_{tot}(0\,\mathrm{kOe})}$ at the same temperature range. The insets to Figs. 3(b), 3(d), 3(f) illustrate the comparison between $\boldsymbol{\kappa_{tot}(0\,\mathrm{kOe})}$ and $\boldsymbol{\kappa_e(0\,\mathrm{kOe})}$.**

## IV. Discussion

The theoretical fit to data in Figs. 2 and 3 using eqns. (7), (9), (10) and (4) accurately describes the dynamic transport response between magnetoresistance, electrical conductivity and total thermal conductivity. The high correlation between the theoretical fit of eqn. (4) to data for the Lorenz ratio indicates the validity of the formula, as well as the different scattering relaxation times in the thermal and electrical processes. It is important to note that when comparing the extrapolation of $\kappa_{ph}$ for 5 K and 40 K in Fig. 2 for all samples to the fit to eqn. (11), error is considerably less at the lower temperature than at the higher temperature. The extrapolation at higher temperature may overestimate $\kappa_{ph}$ due to the fitting parameters of eqn. (11) and can result in a slightly larger ratio of $\kappa_{ph}/\kappa_{tot}$. In order to reduce overestimating the ratio of $\kappa_{ph}/\kappa_{tot}$, this would require an improved theory of electron dominant metals and different functional form of $\kappa_{tot}(\sigma)$ not currently available.

Another method to reduce an overestimation of the ratio of $\kappa_{ph}/\kappa_{tot}$ is to use samples with lowest possible impurity concentration or highest residual resistivity ratio (RRR). Impurities and dislocations can significantly increase electron scattering and proportionally increase the ratio of the phonon contribution. The RRR values of our single crystal metals are given in Table 1 and are on the order of 100, our Zn sample has the highest RRR value and its $\kappa_{ph}$ ratio is also the lowest at low temperatures. As a result, samples with lower impurity and higher RRR values can be considered to reduce the overestimation of $\kappa_{ph}$ ratio [2, 17, 18, 25].

Fig. 4 shows $\kappa_{ph}$ vs. $T$ and illustrates how the different scattering mechanisms in Al, Cu and Zn single crystals change across the temperature spectrum of 5 K - 60 K. From phonon transport theory at low temperatures ($T \ll \Theta_D$), the phonons scatter with electrons resulting in $\kappa_{ph}$ with a temperature dependence of $T^2$ and at even lower temperatures $\kappa_{ph}$ exhibits a $T^3$ law behavior following from the heat capacity of phonons. In the high temperature regime ($T \gg \Theta_D$) the Umklapp process dominates the scattering, with a $T^{-1}$ dependence. The change in the power law behavior as a function of $T$, as shown in all three curves in Fig. 4, indicates the change in the scattering process. This change results in a peak in the $\kappa_{ph}$~$T$ curve, which usually occurs around 10% of the Debye temperature [1, 2] and is demonstrated in all three curves.

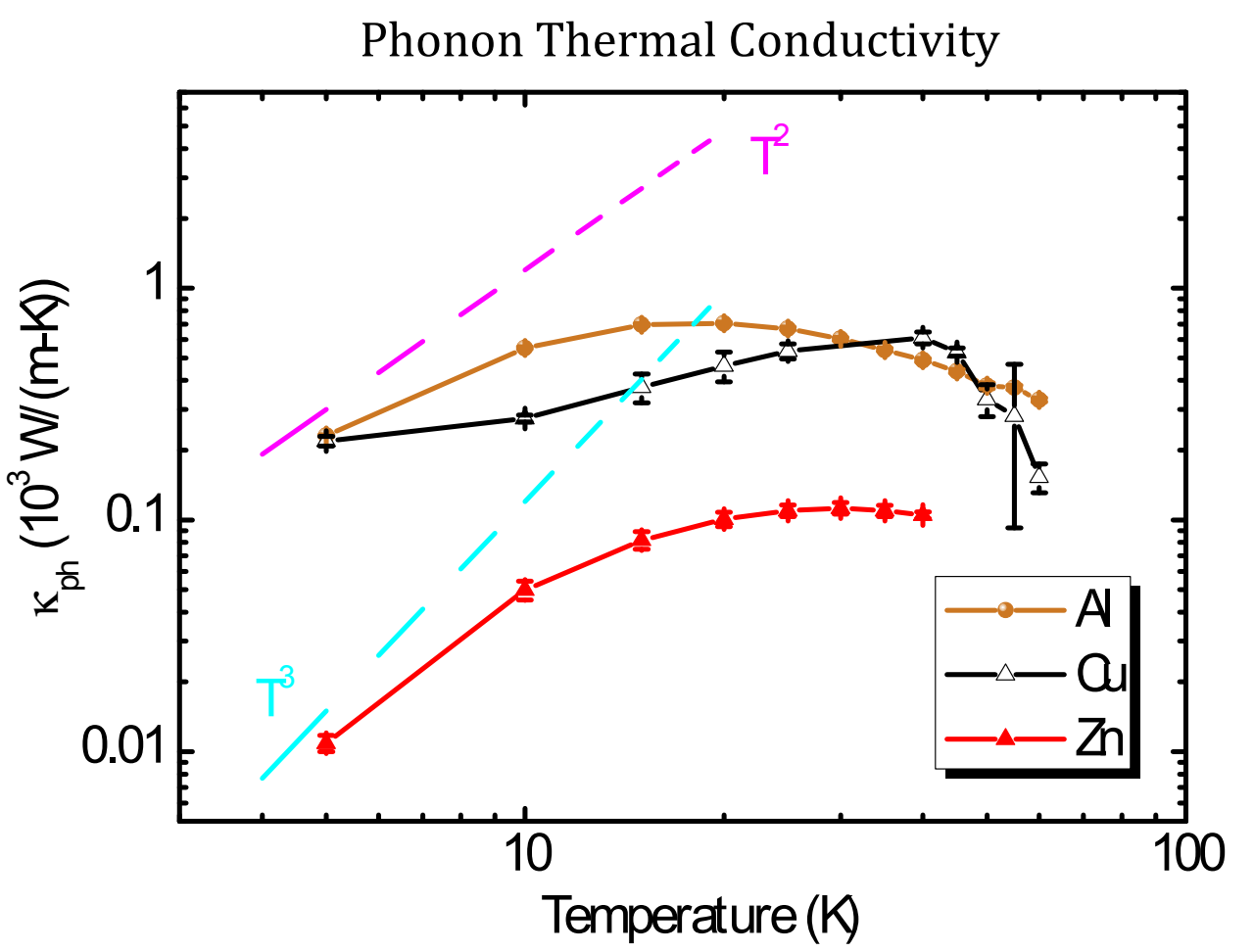

**FIG. 4: Phonon thermal conductivity of Al, Cu and Zn are plotted against temperature in a *log-log* scale. Two dashed lines are a guide to the eye showing different T power law values across the temperature range**.

A requirement of the technique employed here is high sample mobility (*e.g.* $> 10^3$ cm$^2$ V$^{-1}$ s$^{-1}$), so that a large magnetic field (~100 kOe) can suppress the electronic transport. Furthermore, the samples should have a high electrical conductivity, otherwise, the relative change of electrical conduction with the field will be too small for comparison and all the points in the $\kappa_{tot}$~$\sigma$ plot will collapse, making the extrapolation of $\sigma = 0$ unreliable.

In the past, due to the limited availability of large magnetic fields and convenient methods to measure thermal conductivity while applying magnetic fields, very few single crystal metals were explored in transverse field to extract the phonon thermal conductivity or the Lorenz ratio. Some metals, such as Al [18], Cu [6, 19, 20, 25], Zn [21], Au [18], Ag [18], have results for their Lorenz ratio utilizing $\kappa_{tot}$ instead of $\kappa_e$. However, the method employed here lacks the functionality of estimating $\kappa_e$ from $\sigma$. Other research on Cu [5] and Ni [7] uses a diluted alloy approach to extrapolate the behavior of $\kappa_{ph}$ and estimate its elemental value. Our method provides a much simpler means of extracting $\kappa_e$ from $\sigma$. Furthermore, the benefit of measuring single crystals is to make possible a comparison between experimental data and future first principle calculations of the field effect on these transport coefficients. This comparison between theory and experiment is critical information in understanding not only metals, but to better understand a variety of semiconductor materials used in thermoelectric applications.

The method described here can be applied to semiconductor single crystals, such as $Bi_2Te_3$ and $Bi_2Se_3$, which are widely used in thermoelectric applications. A determination of $\kappa_e$ for TE materials is helpful when attempting to reduce $\kappa_{tot}$ in order to improve efficiency. In both of these semiconductors, the phonon contribution to $\kappa_{tot}$ is comparable with the electronic counterpart, or may even be dominant across the temperature spectrum. As seen in metals, the Lorenz ratio drops quickly in the intermediate temperature range, as seen in Fig. 3. If the Sommerfeld value is used to estimate $\kappa_e$, then the real electronic thermal conductivity can be overestimated. The advantage of employing our experimental method on semiconductors is the ability of making Hall measurements on samples to obtain $\rho_{yx}$. Measuring both $\rho_{xx}$ and $\rho_{yx}$ leads to a greater precision in deriving sigma rather than fitting from eqn. (7).

## V. Conclusions

The phonon thermal conductivity and Lorenz ratio of several single crystal metals (Al, Cu and Zn) are extracted from the total thermal conductivity through magnetothermal resistance (MTR) measurements. We find the phonon thermal conductivity has a peak around $\Theta_D/10$, while the Lorenz ratio deviates from the Sommerfeld value in the intermediate temperature range. This technique could prove helpful in evaluating other materials, particularly semiconductors, used in thermoelectric applications, where the suppression of $\kappa_{tot}$ can lead to enhanced thermoelectric efficiency (ZT).

## Acknowledgements

This work was supported by Solid State Solar - Thermal Energy Conversion Center (S3TEC), an Energy Frontier Research Center funded by the U. S. Department of Energy, Office of Science, Office of Basic Energy Science under award number DE-SC0001299. C.P.O. would like to thank Robert D. Farrell, S.J. and Christopher Noyes for helpful comments on the manuscript, and acknowledges financial support from the Trustees of Boston College. M.Y. is grateful to David Broido for helpful discussions and comments on the manuscript. The work at the University of Virginia is supported by the Air Force Young Investigator Award, grant number FA9550-14-1-0316.